\documentclass[journal]{IEEEtran}
\usepackage{graphicx}
\usepackage{subcaption}
\usepackage{hyperref}
\usepackage[labelformat=simple]{subcaption}
\usepackage{float}
\usepackage{tabularx}
\usepackage{multirow}
\usepackage{amsmath,amssymb}
\usepackage{mathtools}

\usepackage[dvipsnames]{xcolor}
\usepackage{tikz}
\usetikzlibrary{positioning}

\newcommand{\jim}{\mathfrak{j}}
\newcommand{\hmat}{\mathbb{H}}
\newcommand{\aod}{\phi^D}
\newcommand{\aode}{\theta^D}
\newcommand{\aodbs}{\varphi^D}
\newcommand{\aodebs}{\vartheta^D}
\newcommand{\aoa}{\phi^A}

\newcommand{\phase}{\psi}
\newcommand{\beamcb}{\mathcal{B}}
\newcommand{\beamcbpred}{B}

\DeclareMathOperator{\Acc}{Acc}
\DeclareMathOperator{\Tx}{Tx}
\DeclareMathOperator{\argmax}{argmax}
\DeclareMathOperator{\RSS}{RSS}

\DeclareMathOperator{\dB}{dB}

\DeclareMathOperator{\SNR}{SNR}

\title{Beam Index Map Prediction in Unseen Environments from Geospatial Data}

\author{Fabian Jaensch, 
Giuseppe Caire \IEEEmembership{Fellow, IEEE},
{Beg{\"u}m Demir \IEEEmembership{Senior Member, IEEE}}
\thanks{F. Jaensch and G. Caire are with the Communications and Information Theory Group, Technische Universit{\"a}t Berlin, 10623 Berlin, Germany.}
\thanks{{B. Demir is with the Remote Sensing Image Analysis Group, Technische Universit{\"a}t Berlin, 10623 Berlin, Germany, and with the BIFOLD - 
Berlin Institute for the Foundations of Learning and Data, 10623 Berlin, Germany.}}
\thanks{Corresponding author: Fabian Jaensch (email: f.jaensch@tu-berlin.de).}
}

\begin{document}
\maketitle


\begin{abstract}
    In 5G, beam training consists of the efficient association of users to beams for a given beamforming codebook used at the base station and the given propagation environment in the cell. 
    We propose a convolutional neural network approach that leverages the position of the base station and geospatial data to predict beam distributions for all user locations simultaneously. 
    Our method generalizes to unseen environments without site-specific training or specialized sensors. 
The results show that it significantly reduces the number of candidate beams considered, thereby improving the efficiency of beam training.
\end{abstract}

\section{Introduction}\label{sec:introduction}
In 5G/6G (massive) MIMO, large antenna arrays allow the direction of signal energy in specific directions via beamforming.
This is particularly important for mmWave and THz communication due to high attenuation, and it also increases data rates and coverage at lower frequencies.
Base stations can cover cells or sectors with a family of beam patterns (referred to as beamforming codebook) that have much higher gain and much narrower angle width than conventional sector-wide beams. 
The efficient association of users to beams is referred to as beam training and is an important aspect of the 5G physical layer. 
Conventional beam sweeping requires testing a large number of beams in order to find the best combination for a strong communication link and causes significant overhead \cite{3gpp}.

Several works have proposed data-driven methods based on machine learning (ML) leveraging additional sensors to reduce the time and signaling overhead required for beam sweeping.
In a fixed environment, the GPS geolocation of the user equipment (UE) can be used to determine the optimal beam \cite{geoloc}.
To account for dynamic changes such as vehicular movement, prior works have used radar, LiDAR or cameras at the base station (BS) or UE for blockage detection and beam selection \cite{dias19,klautau19,wang19}.
Temporal correlation has also been exploited to predict future beam indices from the current optimal choice \cite{tracking} (beam tracking).

The aforementioned methods are site-specific, i.e. they rely on the collection of data for a specific environment and the trained models cannot be used in new environments.
Consequently, in every new environment, data has to be collected and the training procedure needs to be repeated.
In contrast, \cite{tune} investigates applying a model trained in one scenario to another one leveraging transfer learning and incorporating information from cameras, LiDAR and GPS.
Furthermore, \cite{point_cloud_beams} proposes to encode the geometry of the environment in the form of a point cloud derived from stereo imagery, which is then used as an input.
While the resulting model is site-agnostic, it requires multiple photos of the environment from various locations and angles.
One reason for the shortage of site-agnostic methods is presumably the lack of appropriate datasets.
The largest openly available MIMO channel dataset, DeepMIMO(v3) \cite{deepmimo}, contains only 20 city environments with 3 base stations each—far below the hundreds of environments required for training environment-agnostic models in related tasks such as radio map prediction \cite{radiounet}.

Inspired by our recent works on radio map prediction \cite{radiounet,imgopt}, we propose a different point of view on the problem.
In a static environment, the optimal beam index at each UE location is deterministically determined by the environment's geometry and the BS parameters.
We consider the problem of predicting a map of optimal beam indices using ML, based on publicly available city height maps or aerial images, together with properly encoded information about the BS.
Our contributions are as follows:
\begin{itemize}
    \item We formulate the beam index prediction problem as an image-to-image translation task and solve it using convolutional neural networks (CNN). 
        The trained models are not site-specific, but rather learn a mapping from the input geometry to the beam indices, so that they can be used in unseen environments.
    \item Simulation environments and CNN input data are derived from public geospatial databases. 
        Besides the UE location, no further sensor data is required. 
        We release a large synthetic channel dataset with a total of 20151 simulations in 609 city environments, each simulation featuring one BS and tens of thousands of Rx.
        \footnote{Our code is published on \url{https://github.com/fabja19/beam_index_prediction} and the dataset on \url{https://zenodo.org/uploads/17290206}.}
    \item The model predicts a beam candidate map for the entire area of interest in a single forward pass. 
        It can be used by several UE and to update beam indices as UEs move.
    \item We compare the performance of models trained with different loss functions and input data. 
\end{itemize}

\section{System Model and Problem Definition}\label{sec:system}

\subsection{Channel Model}\label{sec:model}
\begin{figure}[!htb]
    \centering
    \tikzset{every node/.style={font=\normalsize}}
    \tikzset{every picture/.style={line width=0.75pt}} 
    \begin{tikzpicture}[x=0.75pt,y=0.75pt,yscale=-1,xscale=1,scale=0.9,every node/.append style={scale=1}]

    \draw  [draw opacity=0] (174.14,59.1) -- (354.53,57.56) -- (300.22,158.64) -- (119.83,160.18) -- cycle ; \draw   (174.14,59.1) -- (119.83,160.18)(194.14,58.93) -- (139.83,160.01)(214.14,58.76) -- (159.83,159.84)(234.14,58.59) -- (179.83,159.67)(254.14,58.42) -- (199.83,159.5)(274.14,58.25) -- (219.83,159.33)(294.14,58.08) -- (239.83,159.15)(314.14,57.91) -- (259.83,158.98)(334.14,57.74) -- (279.82,158.81)(354.13,57.56) -- (299.82,158.64) ; \draw   (174.14,59.1) -- (354.53,57.56)(163.34,79.2) -- (343.73,77.65)(152.55,99.29) -- (332.93,97.75)(141.75,119.38) -- (322.14,117.84)(130.96,139.48) -- (311.34,137.93)(120.16,159.57) -- (300.55,158.03) ; \draw    ;
    \draw [color={rgb, 255:red, 255; green, 0; blue, 0 }  ,draw opacity=1 ]   (236.8,109.6) -- (390.6,110.19) ;
    \draw [shift={(392.6,110.2)}, rotate = 180.22] [color={rgb, 255:red, 255; green, 0; blue, 0 }  ,draw opacity=1 ][line width=0.75]    (10.93,-3.29) .. controls (6.95,-1.4) and (3.31,-0.3) .. (0,0) .. controls (3.31,0.3) and (6.95,1.4) .. (10.93,3.29)   ;
    \draw [color={rgb, 255:red, 255; green, 0; blue, 0 }  ,draw opacity=1 ]   (236.8,109.6) -- (296.02,171.55) ;
    \draw [shift={(297.4,173)}, rotate = 226.29] [color={rgb, 255:red, 255; green, 0; blue, 0 }  ,draw opacity=1 ][line width=0.75]    (10.93,-3.29) .. controls (6.95,-1.4) and (3.31,-0.3) .. (0,0) .. controls (3.31,0.3) and (6.95,1.4) .. (10.93,3.29)   ;
    \draw [color={rgb, 255:red, 255; green, 0; blue, 0 }  ,draw opacity=1 ]   (236.8,109.6) -- (282.47,22.77) ;
    \draw [shift={(283.4,21)}, rotate = 117.74] [color={rgb, 255:red, 255; green, 0; blue, 0 }  ,draw opacity=1 ][line width=0.75]    (10.93,-3.29) .. controls (6.95,-1.4) and (3.31,-0.3) .. (0,0) .. controls (3.31,0.3) and (6.95,1.4) .. (10.93,3.29)   ;
    \draw [color={rgb, 255:red, 74; green, 144; blue, 226 }  ,draw opacity=1 ]   (236.8,109.6) -- (362.97,19.76) ;
    \draw [shift={(364.6,18.6)}, rotate = 144.55] [color={rgb, 255:red, 74; green, 144; blue, 226 }  ,draw opacity=1 ][line width=0.75]    (10.93,-3.29) .. controls (6.95,-1.4) and (3.31,-0.3) .. (0,0) .. controls (3.31,0.3) and (6.95,1.4) .. (10.93,3.29)   ;
    \draw [color={rgb, 255:red, 74; green, 144; blue, 226 }  ,draw opacity=1 ] [dash pattern={on 4.5pt off 4.5pt}]  (236.8,109.6) -- (328.24,92.17) ;
    \draw [shift={(330.2,91.8)}, rotate = 169.21] [color={rgb, 255:red, 74; green, 144; blue, 226 }  ,draw opacity=1 ][line width=0.75]    (10.93,-3.29) .. controls (6.95,-1.4) and (3.31,-0.3) .. (0,0) .. controls (3.31,0.3) and (6.95,1.4) .. (10.93,3.29)   ;
    \draw [color={rgb, 255:red, 74; green, 144; blue, 226 }  ,draw opacity=1 ] [dash pattern={on 4.5pt off 4.5pt}]  (330.2,91.8) -- (363.75,20.41) ;
    \draw [shift={(364.6,18.6)}, rotate = 115.17] [color={rgb, 255:red, 74; green, 144; blue, 226 }  ,draw opacity=1 ][line width=0.75]    (10.93,-3.29) .. controls (6.95,-1.4) and (3.31,-0.3) .. (0,0) .. controls (3.31,0.3) and (6.95,1.4) .. (10.93,3.29)   ;
    \draw  [draw opacity=0] (292.39,99.1) .. controls (293.1,102.55) and (293.54,105.99) .. (293.71,109.4) -- (236.8,109.6) -- cycle ; \draw  [color={rgb, 255:red, 126; green, 211; blue, 33 }  ,draw opacity=1 ] (292.39,99.1) .. controls (293.1,102.55) and (293.54,105.99) .. (293.71,109.4) ;  
    \draw  [draw opacity=0] (288.36,72.89) .. controls (293.59,79.72) and (297.76,87.69) .. (300.47,96.57) .. controls (307.82,120.73) and (302.42,145.41) .. (288.28,161.89) -- (242.49,114.22) -- cycle ; \draw  [color={rgb, 255:red, 126; green, 211; blue, 33 }  ,draw opacity=1 ] (288.36,72.89) .. controls (293.59,79.72) and (297.76,87.69) .. (300.47,96.57) .. controls (307.82,120.73) and (302.42,145.41) .. (288.28,161.89) ;  
    \draw [color={rgb, 255:red, 245; green, 166; blue, 35 }  ,draw opacity=1 ] [dash pattern={on 0.84pt off 2.51pt}]  (330.2,91.8) -- (320.83,109.33) ;
    \draw [color={rgb, 255:red, 245; green, 166; blue, 35 }  ,draw opacity=1 ] [dash pattern={on 0.84pt off 2.51pt}]  (330.2,91.8) -- (246.83,92.33) ;

    \draw (404,101.4) node [anchor=north west][inner sep=0.75pt]   [align=left] {{ x'}};
    \draw (300,166.6) node [anchor=north west][inner sep=0.75pt]   [align=left] {{ z'}};
    \draw (289.2,7.8) node [anchor=north west][inner sep=0.75pt]   [align=left] {{ y'}};
    \draw (283.1,102.8) node  [align=left] {\begin{minipage}[lt]{8.67pt}\setlength\topsep{0pt}
    \textcolor[rgb]{0.49,0.83,0.13}{$\displaystyle \phi $}
    \end{minipage}};
    \draw (293.5,146.8) node  [align=left] {\begin{minipage}[lt]{8.67pt}\setlength\topsep{0pt}
    \textcolor[rgb]{0.49,0.83,0.13}{$\displaystyle \theta $}
    \end{minipage}};
    \draw (322,111) node [anchor=north west][inner sep=0.75pt]   [align=left] {\textcolor[rgb]{0.96,0.65,0.14}{{ dx'}}};
    \draw (215,82) node [anchor=north west][inner sep=0.75pt]   [align=left] {\textcolor[rgb]{0.96,0.65,0.14}{{ dy'}}};

    \end{tikzpicture}

    \caption{AoD in the coordinate system of the array. The intercepts $dx', dy'$  of the projection onto the $x'-y'$-plane with the $x'$- and $y'$-axes correspond to the angles $\varphi, \vartheta$ up to the factor $\pi$ in \eqref{eq:beamspace_angles}.}
    \label{fig:aod_array}
\end{figure}
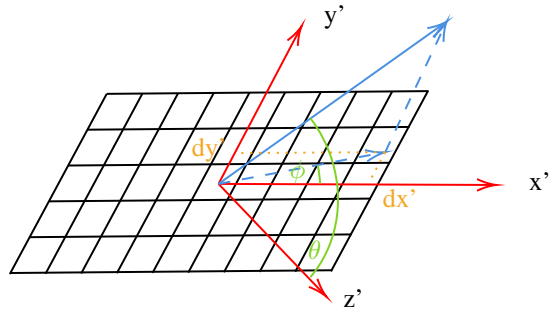

Consider a uniform rectangular array (URA) at a fixed Tx with $N_a,N_e\in\mathbb{N}$ antennas in azimuth and elevation direction, respectively, placed with $\frac{\lambda}{2}$- spacing, where $\lambda > 0$ is the carrier wavelength.
Assuming the geometric multipath channel model \cite{point_cloud_beams}, the downlink channel matrix $\hmat_{\Tx}\in\mathbb{C}^{N_a\times N_e}$ is
\begin{equation}\label{eq:channel_model_tx}
    \hmat_{\Tx}  =   \sum_{p=1}^{N_P} c_p e^{\jim \phase_p} a_{N_a}(\aodbs_p) \otimes a_{N_e}(\aodebs_p),
\end{equation}
where $N_P$ is the number of paths, $c_p > 0$ and $\phase_p\in\left[0, 2\pi\right)$ are the magnitude and the phase of the $p$-th path coefficient, $\jim$ is the imaginary unit, 
\begin{equation}\label{eq:steering_vector}
    a_N(\omega)=\left[1, e^{\jim\omega}, \ldots, e^{\jim\omega(N-1)} \right] \quad\text{ for }N\in\mathbb{N},\,\omega\in\mathbb{R},
\end{equation}
is the steering vector, $\otimes$ denotes the tensor product and $\aodbs_p,\aodebs_p\in [-\pi,\pi]$ are the beamspace angles of departure (AoD).
These are calculated from the AoD $\aod_p,\aode_p$ in a global Cartesian coordinate system as 
\begin{equation}\label{eq:beamspace_angles}
    \begin{split}
        \aodbs_p = \pi  \sin(\aode_p) \cos(\aod_p),\\
        \aodebs_p = \pi \sin(\aode_p) \sin(\aode_p),
    \end{split}
\end{equation}
see Fig. \ref{fig:aod_array}.

On the Rx side, we model the UE as having a small number of coarse angular ``sectors'' (e.g., corresponding to four directional antennas around a smartphone chassis). 
Each arriving path is associated with one such sector according to its azimuth angle of arrival (AoA).
Formally, if we divide the azimuth domain $\left[0,2\pi\right)$ into $N_r$ disjoint sectors, each path with azimuth AoA $\aoa_p$ contributes through a sector-selection vector
\begin{equation}\label{eq:aoa_vector}
    b_N(\aoa_p)\in\left\{0, 1\right\}^{N_r},
\end{equation}
where $b_N(\aoa_p)$ is one-hot at the index of the sector containing $\aoa_p$.
The complete 3-rd order channel tensor in $\mathbb{C}^{N_a\times N_e\times N_r}$ is then 
\begin{equation}\label{eq:channel_model}
    \hmat  =   \sum_{p=1}^{N_P} c_p e^{\jim \phase_p} a_{N_a}(\aodbs_p) \otimes a_{N_e}(\aodebs_p) \otimes b_{N_r}(\aoa_p).
\end{equation}

On the Tx side, we assume a beam codebook consisting of beams generated as tensor products of one-dimensional flat-top beams in both dimensions described in \cite{beamalign}.
We assume here that the size of the Tx codebook is equal to $N_a N_e$, although this is not necessary in general.
The number of beams in each dimension may also be smaller than the number of antennas, depending on the codebook design.
For the Rx, since we are interested in finding the sector with the strongest arriving signal, we chose the canonical basis vectors.
More concretely, $w^H b_{N_r}(\aoa_p)$ is either $0$ or $1$, depending on whether $w$ selects the sector that contains $\aoa_p$ or not.
For brevity, we refer to all elements of the codebook as `beams', although the Rx vectors do not correspond to physically synthesizable beams.
We denote as $\beamcb$ the codebook of size $N_a N_e N_r$ resulting of combining the beams in all dimensions.
The application of $\hmat$ to a beam $u\otimes v \otimes w\in\beamcb$ gives the complex channel gain
\begin{equation}\label{eq:H_beams}
    \begin{split}
        & \mathbb{H} (u\otimes v \otimes w) \\
    = &  \sum_{p=1}^{N_P} c_p e^{\jim \phase_p} u^H a_{N_a}(\aodbs_p) v^Ha_{N_e}(\aodebs_p) w^H b_{N_r}(\aoa_p).
    \end{split}
\end{equation}

\subsection{Objective}\label{sec:objective}
The quality of a link established using a beam $u\otimes v \otimes w\in\beamcb$ is measured in terms of the received signal strength (RSS).
Since in our scenario the wavelength is significantly shorter than the level of detail of the environment data, the phases $\psi_p$ cannot be accurately estimated.
Therefore, we model them as uniformly distributed random variables over $\left[0, 2\pi\right]$ and the RSS in Watts is calculated as
\begin{equation}\label{eq:rss}
    \begin{split}
        & \RSS(u\otimes v\otimes w) =\mathbb{E}\left[| \mathbb{H} (u\otimes v\otimes w)|^2\right]\\ 
            =&  \sum_{p=1}^{N_P} c_p^2 \left|u^H a_{N_a}(\aodbs_p) v^Ha_{N_e}(\aodebs_p) w^H b_{N_r}(\aoa_p)\right|^2.
    \end{split}
\end{equation}
This corresponds to a local time-frequency average of the instantaneous power (see \cite{imgopt} for further discussion).
For a fixed Tx-Rx pair, the \textit{effective channel tensor} is obtained by calculating \eqref{eq:rss} for each beam and collecting the values in a 3-rd order tensor that is an element of $\mathbb{R}^{N_a\times N_e\times N_r}$.

Consider a dataset of $M$ channel samples, each corresponding to one Tx-Rx pair, with RSS functions $\RSS_\ell$, for $\ell=1,\ldots,M$.
Note that in our dataset (Section \ref{sec:dataset}), each channel sample corresponds to a specific environment (city map), a fixed Tx location within that environment, and an Rx location within the Tx's coverage area.
A predictor estimates for each channel sample a probability distribution over the set of possible beams. 
Let
\begin{equation}\label{eq:beam_opt}
    (u\otimes v\otimes w)^*_\ell = \argmax_{u\otimes v\otimes w \in \beamcb} \RSS_\ell(u\otimes v\otimes w)
\end{equation}
be the optimal beam and denote as $\beamcbpred^k_\ell\subseteq\beamcb$ the set of the $k\in\mathbb{N}$ beams with the highest probability according to the predictor.

Ideally, the optimal beam index should coincide with the index that the predictor assigns the highest probability to.
If an exact prediction is infeasible, it is still beneficial to identify a small subset of $k<<N_a N_e N_r$ high-quality candidate beams to reduce the overhead of beam training.
To capture this, we measure the performance of the predictor in terms of top-$k$ accuracy
\begin{equation}\label{eq:topkaccuracy}
    \Acc_k = \frac{1}{M} \sum_{\ell=1}^M \chi_{\beamcbpred^k_\ell}((u\otimes v\otimes w)^*_\ell),
\end{equation}
for $k\in\mathbb{N}, k<N_aN_eN_r$, where $\chi_B$ denotes the indicator function of a set $B$.

However, for the same spatial location, often more than one beam provides a satisfactory signal strength, since different paths (e.g., from distinct reflections) may yield similar signal strength at the same location.
Therefore, as a second evaluation metric, we consider the \textit{throughput ratio (TPR)} \cite{wang19,dias19} defined as 
\begin{equation}\label{eq:atpr}
    T_k = \frac{\sum_{\ell=1}^M \max_{u\otimes v\otimes w \in \beamcbpred^k_\ell} \log_2(1 + \SNR_\ell(u\otimes v\otimes w))}{\sum_{\ell=1}^M \log_2(1 + \SNR_\ell((u\otimes v\otimes w)^*_\ell))}.
\end{equation}
Here, $\SNR_\ell=\frac{\RSS_\ell}{\sigma^2}$ is the signal-to-noise ratio with noise variance $\sigma^2$, which in turn depends on the signal bandwidth and the noise power spectral density (PSD).

\section{Beam Index Map Prediction with CNN}\label{sec:prediction}
Our goal is to predict optimal beam indices—or candidate sets—for all locations in a given city environment using a CNN.
We formulate the task as image-to-image translation, mapping environment and Tx information to a spatial distribution over beam indices.

\subsection{Dataset and Processing}\label{sec:dataset}
\begin{figure}[h]
    \begin{subfigure}{0.13\textwidth}
        \includegraphics[width=\textwidth]{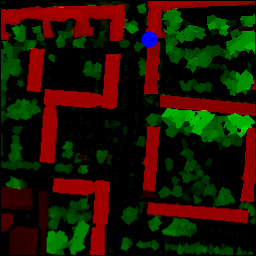}
        \caption{}
        \label{fig:data:city_map}
    \end{subfigure}
    \begin{subfigure}{0.13\textwidth}
        \includegraphics[width=\textwidth]{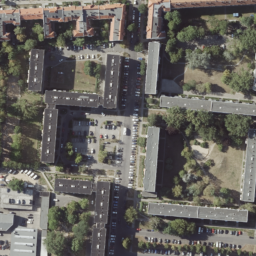}
        \caption{ }
        \label{fig:data:img}
    \end{subfigure}
    \begin{subfigure}{0.13\textwidth}
        \includegraphics[width=\textwidth]{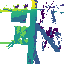}
        \caption{ }
        \label{fig:data:gt}
    \end{subfigure}
    \begin{subfigure}{0.028\textwidth}
        \includegraphics[width=\textwidth]{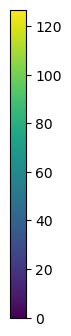}
        \caption*{}
    \end{subfigure}
    \\
    \begin{subfigure}{0.13\textwidth}
        \includegraphics[width=\textwidth]{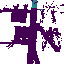}
        \caption{ }
        \label{fig:data:rm1}
    \end{subfigure}
    \begin{subfigure}{0.13\textwidth}
        \includegraphics[width=\textwidth]{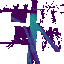}
        \caption{ }
        \label{fig:data:rm2}
    \end{subfigure}
    \begin{subfigure}{0.13\textwidth}
        \includegraphics[width=\textwidth]{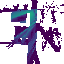}
        \caption{ }
        \label{fig:data:rm3}
    \end{subfigure}
    \begin{subfigure}{0.034\textwidth}
        \includegraphics[width=\textwidth]{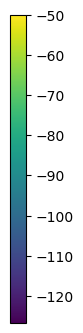}
        \caption*{}
    \end{subfigure}
    \caption{
        (\subref{fig:data:city_map}) City map with overlayed nDSMs of buildings in red, vegetation in green, color intensity correlating with height, and Tx position in blue,
        (\subref{fig:data:img}) aerial image,
        (\subref{fig:data:gt}) ground truth optimal beam indices,
        (\subref{fig:data:rm1})-(\subref{fig:data:rm3}) radio maps depicting the RSS per UE location for three different beams
    }
    \label{fig:data}
\end{figure}

We use our 3D model of Berlin described in \cite{imgopt}, based on open LiDAR data with corresponding aerial images.
Environment information is encoded in \textit{normalized digital surface models (nDSMs)} for the building and vegetation classes, depicting the height of objects per pixel.
Each 256$\times$256 nDSM corresponds to a 256 m$^2$ region at 1m resolution, see Fig. \ref{fig:data} for an example.
For each Tx (placed on rooftop edges) and Rx (on a grid of 1m resolution), both with isotropic antennas, we simulate full path data (coefficients, delays, AoD, AoA) using the ray-tracing software Wireless InSite \cite{wi}.
In total, 20151 simulations have been carried out in 609 city environments at a carrier frequency of $3.9$GHz.

For each grid location (pixel), the effective channel tensor is obtained from the path data as described in Section \ref{sec:system}, with parameters $N_a=8, N_e=N_r=4$.
The orientation of the arrays is chosen so that in azimuth direction they point in straight line away from the building where the Tx is located and the boresight is tilted downwards at a 45$^\circ$ angle from the horizontal plane passing through the Tx.
Postprocessing revealed that the 1$\times$1m spatial resolution combined with the channel tensor dimensions incurred excessive memory usage, implying impractical memory requirements during training.
To resolve this issue, we downscale beam index maps and channel tensors by a factor of 4 in each spatial dimension.
\footnote{To assess the impact of downscaling, we additionally generated high-resolution beam index maps for a small subset of samples.
Interpreting the ranking of beams in each pixel of the downscaled effective channel tensor as an estimate for the corresponding $4\times4$ block in the high-resolution map, we observed a top-1 accuracy of $0.72$ and a top-1 TPR of $0.88$, when referring to the values in the high-resolution map as the ground truth.}

\subsection{Deep Learning Approach}\label{sec:learning}
Tx position and orientation are encoded using one-hot maps, distance maps, and azimuth maps as in \cite{imgopt}.
We also adopt the UNetDCN architecture from there with $C=32,d=5$, previously shown effective for learning radio maps.

This is a semantic segmentation task with the particularity that near-optimal classes may still yield high signal quality.
To account for this, we train the model with several different objective loss functions in separate experiments and compare the performance.
In all cases, we use standard supervised learning with backpropagation.
The number of channels of the output tensor and the last activation function are adapted to the requirements of the loss function.
We evaluate both joint classification over $N_a N_e N_r$ beam indices and separate classification over azimuth, elevation, and Rx sectors (\textit{sep}), when applicable.

We consider the following loss functions:
\begin{itemize}
    \item \textit{Cross Entropy (CE)}: 
        The standard loss function for classification and segmentation problems. 
    \item \textit{Cross Entropy with Probabilities (CEP)}: 
        The effective channel tensor in $\dB$ is normalized and used as a class probability distribution for CE, similar to label smoothing \cite{smoothing}.
    \item \textit{Wasserstein (WS)}: 
        Following the intuition that predicting an adjacent beam to the optimal one should result in lower loss than predicting a beam that is further away, we define beam distances via Euclidean distance between index tuples.
        We calculate the Wasserstein loss \cite{wasserstein} between the probability distribution defined by the output logits and the one-hot target distribution of the optimal index using these distances. 
    \item \textit{Index Regression (IR)}: 
        Prediction of the optimal index is considered as a regression problem, following again the intuition that subsequent beam indices correspond to illuminated areas which are spatially close. 
        This intuition is meaningful only if we consider azimuth, elevation and Rx beams separately. 
    \item \textit{Gain Regression (GR)}: 
        The model is trained to predict the effective channel tensor in dB scale with MSE loss.
 The predictions are ranked according to the order of the estimated RSS values.
\end{itemize}

\section{Results}\label{sec:results}
\subsection{Experimental Setup}\label{sec:results:setup}
We use about 80\% of the dataset for training, 10\% for validation (learning rate reduction, early stopping and selection of the best model state) and 10\% for testing.
Locations with path gain below $-147\dB$ across all beams are excluded, as they imply $\SNR<0$ and are irrelevant for communication.
This is discussed in more detail in \cite{radiounet}, \cite{imgopt}.
This threshold assumes 10 MHz bandwidth, -174 dBm/Hz noise PSD, 23dBm transmit power, and 0dB noise figure, following \cite{imgopt}.

\subsection{Numerical Results on Loss Functions}\label{sec:results:loss_fn}
\begin{table}
    \centering
    \begin{tabular}{|l|cccccc|}
        \hline
        Model & Top-1 & Top-2 & Top-4 & Top-8 & Top-16 & Top-32 \\ \hline
        CE      & \textbf{0.611}   & \textbf{0.736}   & \textbf{0.834}   & \textbf{0.909}   & \textbf{0.963}   & \textbf{0.992} \\ 
        CE sep  & \textit{0.583}   & \textit{0.701}   & \textit{0.784}   & \textit{0.852}   & \underline{0.909} & \underline{0.956} \\ 
        CEP     & 0.431             & 0.614            & \underline{0.748}& \underline{0.844} & \textit{0.921}  & \textit{0.971} \\ 
        CEP sep & 0.461            & \underline{0.619}& 0.728            & 0.813            & 0.893            & 0.951 \\ 
        WS      & 0.347            & 0.412            & 0.459            & 0.506            & 0.547            & 0.591 \\ 
        WS sep  & 0.412            & 0.497            & 0.550            & 0.603            & 0.672            & 0.748 \\ 
        IR sep  & 0.429            & 0.431            & 0.443            & 0.457            & -                 & - \\ 
        GR      & 0.400            & 0.539            & 0.637            & 0.713            & 0.800            & 0.887 \\ 
        GR sep  & \underline{0.484} & 0.616           & 0.699            & 0.775            & 0.855            & 0.919 \\ 
        \hline
    \end{tabular}
    \caption{Top-$k$ accuracy. Best results for each $k$ highlighted in \textbf{boldface}, \textit{italics} and \underline{underlined}.}
    \label{tab:accuracy}
\end{table}

\begin{table}
    \centering
    \begin{tabular}{|l|cccccc|}
        \hline
        Model & Top-1 & Top-2 & Top-4 & Top-8 & Top-16 & Top-32 \\ \hline
        CE      & \textbf{0.974}   & \textbf{0.987}   & \textbf{0.994}   & \textbf{0.998}   & \textbf{0.999}   & \textbf{1.000} \\ 
        CE sep  & \textit{0.963}   & \textit{0.980}   & \textit{0.989}   & \textit{0.993}   & \underline{0.996} & \underline{1.000} \\ 
        CEP     & 0.789            & 0.904            & 0.968            & \underline{0.990} & \textit{0.997}  & \textit{0.999} \\ 
        CEP sep & 0.875            & 0.941            & 0.970            & 0.983            & 0.993            & 0.998 \\ 
        WS      & 0.686            & 0.730            & 0.749            & 0.758            & 0.767            & 0.789 \\ 
        WS sep  & 0.888            & 0.932            & 0.954            & 0.964            & 0.972            & 0.983 \\ 
        IR sep  & 0.862            & 0.870            & 0.879            & 0.882            & - & - \\
        GR      & 0.866            & 0.929            & 0.967            & 0.981            & 0.991            & 0.997 \\ 
        GR sep  & \underline{0.917}& \underline{0.960} & \underline{0.9774} & 0.988          & 0.994           & 0.998 \\ 
        \hline
    \end{tabular}
    \caption{Top-$k$ TPR. Best results for each $k$ highlighted in \textbf{boldface}, \textit{italics} and \underline{underlined}.}
    \label{tab:tpr}
\end{table}

\begin{figure}[h]
    \includegraphics{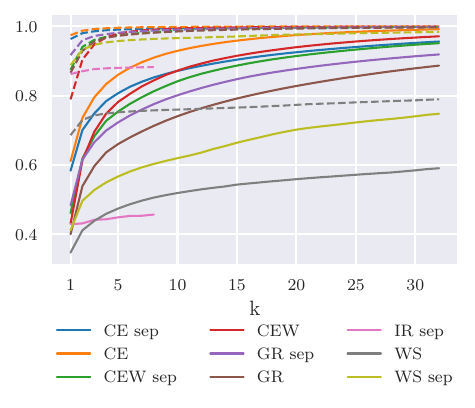}
    \caption{Top-$k$ accuracy (solid) and TPR (dashed) plotted vs $k$}
    \label{fig:curves_acc_tpr}
\end{figure}

The performance of models trained with the loss functions described in \ref{sec:learning} is compared in Tables \ref{tab:accuracy}, \ref{tab:tpr} and the corresponding curves are shown in Fig. \ref{fig:curves_acc_tpr}.  
The results show that standard CE loss over all beam combinations yields the best performance in terms of both evaluation metrics considered, and for all $k$.
From the curves for WS and IR sep in Fig. \ref{fig:curves_acc_tpr}, we observe that a higher accuracy does not necessarily imply a higher TPR as well. 
A high TPR of over $0.99$ is already achieved for $k=4$, although a high accuracy of at least $0.9$ or $0.99$ requires $k=8$ or $k=32$ beams, respectively.
This indicates that suboptimal beams often yield comparable RSS, due to overlapping beam footprints, reflections, and sidelobes — consistent with findings in \cite{wang19,tracking}.

\begin{figure}[h]
    \includegraphics{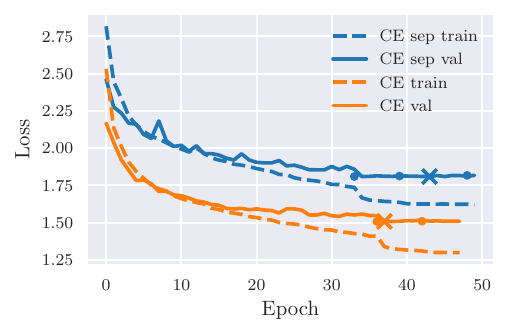}
    \caption{Training and validation losses for CE. Dots indicate learning rate reduction; the cross marks the best epoch in terms of validation loss.}
    \label{fig:curves_losses}
\end{figure}

\begin{figure*}[h]
    \centering
    \includegraphics[width=0.11\textwidth]{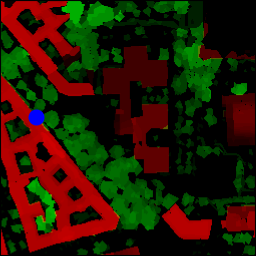}
    \includegraphics[width=0.11\textwidth]{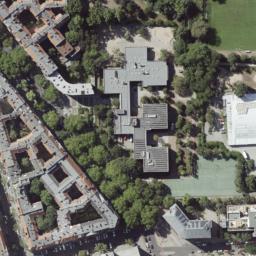}
    \includegraphics[width=0.11\textwidth]{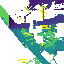}
    \includegraphics[width=0.023\textwidth]{colorbar}
    \includegraphics[width=0.11\textwidth]{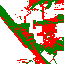}
    \includegraphics[width=0.11\textwidth]{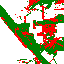}
    \includegraphics[width=0.11\textwidth]{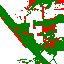}
    \includegraphics[width=0.11\textwidth]{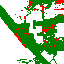}
    \includegraphics[width=0.11\textwidth]{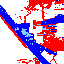}
    \vspace{0.11cm}
    \\
    \includegraphics[width=0.11\textwidth]{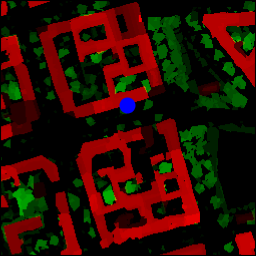}
    \includegraphics[width=0.11\textwidth]{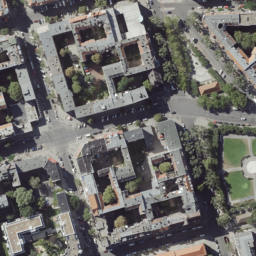}
    \includegraphics[width=0.11\textwidth]{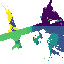}
    \includegraphics[width=0.023\textwidth]{colorbar}
    \includegraphics[width=0.11\textwidth]{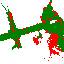}
    \includegraphics[width=0.11\textwidth]{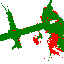}
    \includegraphics[width=0.11\textwidth]{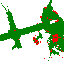}
    \includegraphics[width=0.11\textwidth]{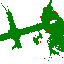}
    \includegraphics[width=0.11\textwidth]{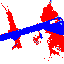}
    \vspace{0.11cm}
    \\
    \begin{subfigure}{0.11\textwidth}
        \includegraphics[width=\textwidth]{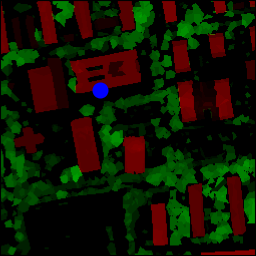}
        \caption{}
        \label{fig:pred_city_map}
    \end{subfigure}
    \begin{subfigure}{0.11\textwidth}
        \includegraphics[width=\textwidth]{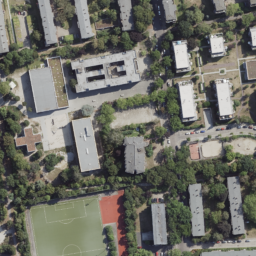}
        \caption{}
        \label{fig:pred_img}
    \end{subfigure}
    \begin{subfigure}{0.11\textwidth}
        \includegraphics[width=\textwidth]{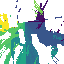}
        \caption{}
        \label{fig:pred_gt}
    \end{subfigure}
    \begin{subfigure}{0.023\textwidth}
        \includegraphics[width=\textwidth]{colorbar}
        \caption*{}
    \end{subfigure}
    \begin{subfigure}{0.11\textwidth}
        \includegraphics[width=\textwidth]{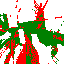}
        \caption{}
        \label{fig:pred_top1}
    \end{subfigure}
    \begin{subfigure}{0.11\textwidth}
        \includegraphics[width=\textwidth]{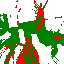}
        \caption{}
        \label{fig:pred_top2}
    \end{subfigure}
    \begin{subfigure}{0.11\textwidth}
        \includegraphics[width=\textwidth]{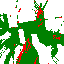}
        \caption{}
        \label{fig:pred_top4}
    \end{subfigure}
    \begin{subfigure}{0.11\textwidth}
        \includegraphics[width=\textwidth]{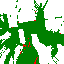}
        \caption{}
        \label{fig:pred_top8}
    \end{subfigure}
    \begin{subfigure}{0.11\textwidth}
        \includegraphics[width=\textwidth]{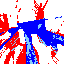}
        \caption{}
        \label{fig:pred_los}
    \end{subfigure}
    \vspace{0.11cm}
    \caption{
        (\subref{fig:pred_city_map}) City map (explained in Fig. \ref{fig:data}),
        (\subref{fig:pred_img}) aerial image,
        (\subref{fig:pred_gt}) ground truth optimal beam indices,
        (\subref{fig:pred_top1})-(\subref{fig:pred_top8}) areas where the optimal beam index is contained in the set of top-$k$ predicted beams in green, others in red, for $k=1,2,4,8$,
        (\subref{fig:pred_los}) areas with a straight path from the Tx in blue, darker blue if this is the strongest among all arriving paths.
    }
    \label{fig:pred}
\end{figure*}

In Fig. \ref{fig:curves_losses}, we show the loss curves during training and Fig. \ref{fig:pred} provides some samples from the dataset together with the areas in which the top-$k$ predicted beams contain the optimal one.
Here, we used the model trained with CE.
The last column also shows the areas with line-of-sight (LoS) to the Tx.
LoS areas (dominant path is direct) are shown in dark blue areas with attenuated LoS due to trees in light blue, areas without a direct path in red.
Performance is strongest in LoS regions, as expected, but high top-$4$/top-$8$ coverage is also achieved in NLoS regions.

\subsection{Numerical Results on Input Data}\label{sec:results:inputs}
Following \cite{imgopt}, we retrain models on either aerial imagery only or imagery with an unclassified nDSM (combined height of buildings and vegetation) using CE loss.
These inputs are more widely available, broadening applicability.
The results in Table \ref{tab:img:accuracy} and \ref{tab:img:tpr} show that also without explicit access to the classes and even without height information, the model retains high prediction accuracy.

\begin{table}
    \centering
    \begin{tabular}{|l|cccccc|}
        \hline
        Inputs      & Top-1             & Top-2             & Top-4            & Top-8              & Top-16            & Top-32 \\ \hline
        Baseline    & {0.611}           & {0.736}           & {0.834}           & {0.909}           & {0.963}           & {0.992} \\ 
        Image& 0.586             & 0.706             & 0.804            & 0.888              & 0.952            & 0.989 \\
        Image+nDSM& 0.606        & 0.733             & 0.832            & 0.909              & 0.962             & 0.991 \\ 
        \hline
    \end{tabular}
    \caption{Top-$k$ accuracy for modified inputs. Baseline is CE in Table \ref{tab:accuracy}.}
    \label{tab:img:accuracy}
\end{table}

\begin{table}
    \centering
    \begin{tabular}{|l|cccccc|}
        \hline
        Inputs          & Top-1 & Top-2 & Top-4 & Top-8 & Top-16 & Top-32 \\ \hline
        Baseline        & {0.974}       & {0.987}           & {0.994}           & {0.998}           & {0.999}           & {1.000} \\ 
        Image           & 0.970         & 0.984             & 0.992            & 0.997              & 0.999            & 1.000 \\
        Image + nDSM    & 0.972        & 0.986             & 0.994            & 0.997              & 0.999             & 1.000 \\
        \hline
    \end{tabular}
    \caption{Top-$k$ TPR for modified inputs. Baseline is CE in Table \ref{tab:tpr}.}
    \label{tab:img:tpr}
\end{table}

\section{Conclusion}\label{sec:conclusion}
We demonstrate that CNN-based models can significantly reduce beam training overhead using only public geospatial data and Tx/UE positions.
As the model predicts beam indices for all locations, inference is required only once per area, not per UE position.
This also enables multi-user applicability from a single inference pass.

One limitation of our experimental setup is that all environments are static.
For future work, it would be interesting to incorporate changes in the environment such as cars moving along the streets.
Since inference with the trained models is very fast, the beam index map could be updated periodically.

While our method generalizes well across places in the same city, we expect it to be less accurate in cities with a very different distribution of buildings and vegetation.
Retraining on a broader dataset spanning multiple cities would likely mitigate this limitation.

\bibliographystyle{IEEEtran}
\bibliography{bib_beams}

\end{document}